\documentclass[prd,twocolumn,superscriptaddress,floatfix,amsmath,amssymb]{revtex4-1}
\usepackage[english]{babel}
\usepackage{graphicx}
\usepackage{dcolumn}
\usepackage{bm}
\usepackage{verbatim}
\usepackage{dsfont}
\usepackage{cancel}
\usepackage{enumerate}
\usepackage{color}
\usepackage{hyperref}
\usepackage{verbatim}
\usepackage{mathrsfs}
\usepackage{hyperref}	
\usepackage{cancel}
\usepackage{bm}
\usepackage{braket}
\usepackage{slashed}
\newcommand{\proj}[2]{\left| {#1} \right\rangle\!\left\langle {#2} \right|}


\begin{document}

\title{%
	Precise space-time positioning for entanglement harvesting%
	}

\author{Eduardo Mart\'{i}n-Mart\'{i}nez}
\affiliation{%
	Institute for Quantum Computing, University of Waterloo, Waterloo, Ontario, N2L 3G1, Canada%
	}
\affiliation{Department of Applied Mathematics, University of Waterloo, Waterloo, Ontario, N2L 3G1, Canada}
\affiliation{Perimeter Institute for Theoretical Physics, 31 Caroline St N, Waterloo, Ontario, N2L 2Y5, Canada}

\author{Barry C. Sanders}
\affiliation{%
	Institute for Quantum Computing, University of Waterloo, Waterloo, Ontario, N2L 3G1, Canada%
	}
\affiliation{%
    Institute for Quantum Science and Technology, University of Calgary, Alberta, Canada T2N 1N4%
    }
\affiliation{%
    Program in Quantum Information Science,
    Canadian Institute for Advanced Research,
    Toronto, Ontario M5G 1Z8, Canada%
    }
\affiliation{%
    Hefei National Laboratory for Physical Sciences at Microscale and Department of Modern Physics,
    University of Science and Technology of China, Hefei, Anhui 230026, China
    }
\affiliation{%
    Shanghai Branch,
    CAS Center for Excellence and Synergetic Innovation Center
        in Quantum Information and Quantum Physics,
    University of Science and Technology of China, Shanghai 201315, China
    }

\begin{abstract}
We explore the crucial role of relative space-time positioning between the two detectors in an operational two-party entanglement-harvesting protocol.
Specifically we show that the protocol is robust if imprecision in spatial positioning and clock synchronization are much smaller than the spatial separation between the detectors and its light-crossing time thereof. This in principle guarantees robustness if the imprecision is comparable to a few times the size of the detectors, which suggests entanglement harvesting could be explored for tabletop experiments. On the other hand, keeping the effects of this imprecision under control would be demanding on astronomical scales.
\end{abstract}

\maketitle

\section{Introduction}
\label{sec:intro}

The vacuum state of a long-range field is unentangled with respect to its extended nonlocal modes
but can be entangled with respect to local modes~\cite{Unr76}
corresponding to localized detectors. 
Extracting this naturally endowed 
resource from a quantum field vacuum via localized detectors,
such as atomic probes,
is called \emph{vacuum entanglement harvesting}~\cite{Valentini1991,Rez03,RRS05,Salton:2014jaa,Pozas2015}.
The process can be understood in the following terms:
An actuator turns the interaction between these two detectors and the field on and off,
resulting in two initially unentangled detectors evolving into an entangled pair,
after ignoring, or tracing out, the field itself.
This process is valid even for two spacelike-separated detectors.

The resultant pair of entangled detectors
can in principle serve as a resource for performing quantum information tasks
such as teleportation~\cite{BBC+93,BPM+97},
superdense coding~\cite{BW92,MWKZ96},
or fingerprinting~\cite{BCWdW01,HBM+05}.
This resource is especially useful when enhanced by
entanglement-farming protocols using successive pairs of detectors~\cite{MBDK13}.
We are particularly interested in the role and limitations to this protocol that arise form the requirement to control the spatial separation between the detectors and their switching synchronization.

Previous studies do not consider the spacetime position problem
but rather assume that separation is known and a precise clock
is shared between the two parties
(henceforth Alice or A and Bob or B)
undertaking the entanglement-harvesting protocol.
Here we establish the sensitivity of entanglement harvesting
under imprecision in spacetime positioning and determine when the protocol is robust under such imprecision.

The usual model for entanglement harvesting treats two localized
Unruh-DeWitt detectors~\cite{Unr76,DeW79} linearly
coupled to a scalar field in a vacuum state according to well behaved switching functions~\cite{LS06,LS08}. Despite its simplicity, the Unruh-DeWitt model successfully
captures essential features of the light-matter interaction~\cite{Wavepackets,Alvaro}. We treat the field as occupying a $(3+1)$-dimensional flat geometry 
with spacetime coordinate \mbox{$x:=(\bm{x},t)$}.
Bipartite entanglement can be assessed by calculating negativity~\cite{Negat}. A positive value of negativity is a necessary and sufficient condition for entanglement in the case of two qubits~\cite{ZHSL98,Ple05}.


\section{Entanglement harvesting protocol}
\label{sec:protocol}
An Unruh-DeWitt detector is a localized particle centered at~$\bm{x}$
with spatial smearing function $F(\bm x)$
(e.g., spatial wavefunction of electron for single-atom detector)
and has two distinct internal energy levels~$\ket{g}$ and~$\ket{e}$ for ground and excited state,
respectively, separated by an energy gap~$\Omega$.
For $\ket{e}=\hat{\sigma}^+\ket{g}$,
working in the interaction picture, the detector's monopole moment is~\cite{Alvaro,Wavepackets}
\begin{equation}
\label{eq:detectormoment}
	\hat{\mu}(t)
		=\hat{\sigma}^+\text{e}^{\mathrm{i}\Omega t}
			+\hat{\sigma}^-\text{e}^{-\mathrm{i}\Omega t}.
\end{equation}
The detector monopole's moment is linearly coupled to a massless scalar field,
whose operator in the interaction picture is given by
\begin{equation}
	\hat{\phi}(x)
		=\int\!\frac{\text{d}^3\bm k}{\sqrt{(2\pi)^32\omega}}
			\left[\hat{a}\!\left({\bm k}\right) \text{e}^{-\mathrm{i}k\cdot x}+\text{H.c.}\right]
\end{equation}
for $k:=(\bm{k},\omega=\left|\bm{k}\right|)$ with H.c. designating Hermitian conjugate.
We denote the $\bm{k}^\text{th}$-mode annihilation operator by
$\hat{a}\!\left({\bm k}\right)$,
and its adjoint creates the plane-wave single-photon Fock state
\begin{equation}
	\hat{a}^\dagger(\bm{k})\ket{0}=\ket{1(\bm{k})}
\end{equation}
from the vacuum~$\ket{0}$.

Given field-detector coupling strength~$\lambda$ and switching function~$\chi(t)$,
the interaction Hamiltonian is
\begin{equation}
	\hat{H}_\text{I}=\lambda\chi(t)\hat{\mu}(t)\!\int\!\text{d}^3\,\bm{x}F(\bm x)\hat{\phi}(\bm x,t),
\end{equation}
where $F(\bm x)$ is the spatial profile of the detector. For a non-pointlike profile,  the behaviour of the detector is regular even for piecewise continuous switching functions~\cite{LS06}.  We introduce the detector's form factor
\begin{equation}
\tilde F_\nu(\bm k)
	=\int\!\text{d}^3 \bm x\,(2\pi)^{{-}\frac{3}{2}} F_\nu(\bm x)\text{e}^{\mathrm{i}\bm k\cdot \bm x},
\end{equation}
where $\nu\in\{\text{A,B}\}$ labels Alice's and Bob's detectors. From this, we write
\begin{align}
	\hat{H}_\text{I}\!
		=&\sum_{\nu=A,B}\lambda_\nu\chi_\nu(t)\hat{\mu}_\nu(t)
					\nonumber\\&\times
			\int\frac{\text{d}^3\bm  k}{\sqrt{(2\pi)^32\omega}}
			\left(\text{e}^{-\mathrm{i}k\cdot x_\nu}\hat{a}\left({\bm k}
				\right)\tilde F_\nu(\bm k)+\text{H.c.}\right),
\end{align}
and the perturbative evolution operator is
\begin{equation}
	\openone\underbrace{-\mathrm{i}\int_{-\infty}^{\infty}\text{d}t\hat{H}_\text{I}(t)}_{U^{(1)}}\underbrace{ - \int_{-\infty}^{\infty}\text{d}t \int_{-\infty}^{t}\text{d}t'\hat{H}_\text{I}(t) \hat{H}_\text{I}(t')}_{U^{(2)}}+O\left(\lambda^3\right)
\end{equation}
for
\begin{equation}
	\lambda:=\max_\nu\lambda_\nu\ll\Omega:=\min_\nu\Omega_\nu.
\end{equation}

\begin{figure}
	\includegraphics[width=.80\columnwidth]{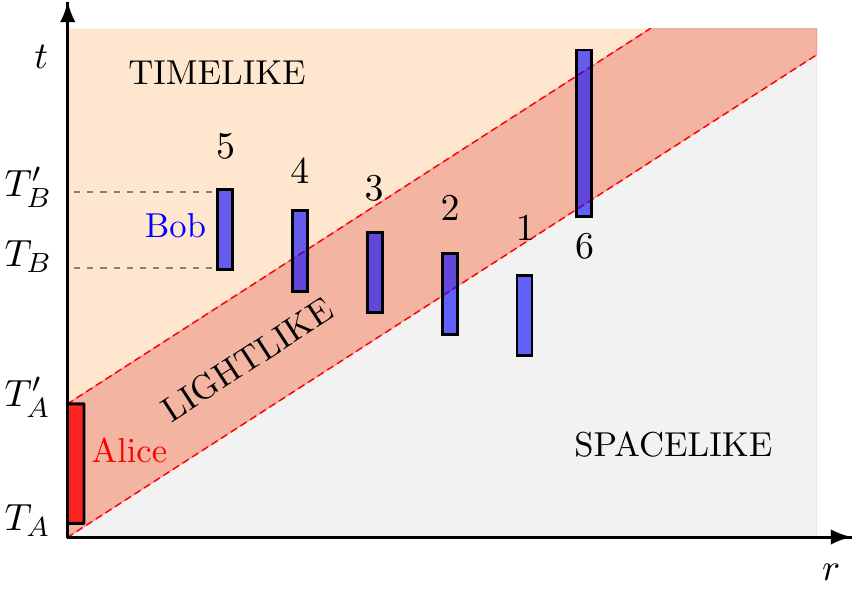}\\
		\caption{%
	(Color online)
	Spacetime diagram showing the possible relative placement of the switching periods of Alice's and Bob's detectors. We follow the same conventions as in~\cite{Blasco2015,Blasco2016}, showing the following possible relative position of the detectors Alice and Bob: 1-purely spacelike, 2,3,4-(partially) light connected, 5-purely timelike, 6-lightlike only for part of Bob's interaction.}
	\label{diagram}
\end{figure}

Inter-detector separation and detector switching can be partitioned into two regimes:
lightlike contact,
whereby real quanta can be exchanged,
and spacelike separation,
whereby they cannot be exchanged.
Notice that it is possible that the two detectors communicate  when they are in timelike contact
even though energy does not necessarily flow from Alice to Bob~\cite{Comm2,Blasco2015,EMM2015,JonssonS}.
In the spacetime diagram in Fig.~\ref{diagram},
we present the possible cases of relative positioning of Alice's and Bob's detectors, following the same conventions as in~\cite{Blasco2015,Blasco2016}.

The final state for the detectors and field is
\begin{equation}
	\rho=\rho_0+\rho^{(1)}+\rho^{(2)}+O(\lambda^3)
\end{equation}
for initial ground state
\begin{equation}
	\rho_0:=\ket{0gg}\bra{0gg}
\end{equation}
and
\begin{align}
	\rho^{(1)}=&U^{(1)}\rho_0+\rho_0{U^{(1)}}^\dagger,\,
	\rho^{(2)}=\rho_1^{(2)}+\rho_2^{(2)},\nonumber\\
	\rho_1^{(2)}:=&U^{(1)}\rho_0{U^{(1)}}^\dagger,
	\rho_2^{(2)}:=U^{(2)}\rho_0+\rho_0{U^{(2)}}^\dagger.
\end{align}
The final reduced two-detector state,
\begin{equation}
	\rho_\text{AB}:=\operatorname{tr}_\text{f}\rho
\end{equation}
for~tr$_\text{f}$ denoting partial trace over the field,
{satisfies}
\begin{equation}
	\operatorname{tr}_{\text{f}}\left(U^{(1)}\rho_0\right)=0
\end{equation}
{so} the leading-order change to~$\rho_\text{AB}$ is~$O(\lambda^2)$.

Let~Alice and Bob employ identical detectors:
$F(\bm x)$ and $\lambda$ are the same.
We assume a real even spatial profile $F(\bm x)= F(-\bm x)$, so
\begin{equation}
	\tilde F(\bm k)=\tilde F(-\bm k).
\end{equation}
{We i}ntroduc{e} shifted frequencies
\begin{equation}
	\omega_\nu^\pm(\bm{k}):=\omega(\bm{k})\pm\Omega_\nu
\end{equation}
and integrals
\begin{equation}
\label{eq:Inuintegral}
	I_{\nu}(\bm k)
		:=\lambda_\nu\frac{\text{e}^{-\mathrm{i} \bm k\cdot\bm x_\nu}}{\sqrt{ 2\omega}}
			\tilde{F}(\bm k)
			\int_{-\infty}^\infty\!\!\! \text{d}t\,\chi_\nu(t)\,\text{e}^{\mathrm{i}\omega_\nu^+(\bm{k})t}
\end{equation}
and,
for $\bm{r}:=\bm{x}_\text{A}-\bm{x}_\text{B}$
(and $r:=\left|\bm{r}\right|$),
\begin{align}
\label{eq:Jintegral}
	J(\bm k)
		:=&\lambda_\text{A}\lambda_\text{B}\int_{-\infty}^{\infty}\text{d}t\int_{-\infty}^{t}\text{d}t'
			\Big[\chi_\text{A}(t')\chi_\text{B}(t)
				\nonumber\\&\times
			\text{e}^{\mathrm{i}\left(\omega_{\text A}^+(\bm{k}) t'-\omega_{\text B}^-\left(\bm{k}\right)t\right)}
				\nonumber\\&+
			\chi_\text{A}(t)\chi_\text{B}(t')\text{e}^{\mathrm{i}\left(\omega_{\text B}^+(\bm{k})t'-\omega_{\text A}^-\left(\bm{k}\right)t\right)}\Big]
		\frac{\tilde{F}(\bm k)^2}{2\omega }{\text{e}^{\mathrm{i} \bm k\cdot\bm{r}}}.
\end{align}
{Then w}e define
\begin{equation}
\label{eq:IJtotal}
	\mathcal{I}_{\nu\nu'}:=\int\text{d}^3\,\bm{k}I^*_\nu(\bm{k})I_{\nu'}(\bm{k}),\;
	\mathcal{J}:=\int\text{d}^3\bm{k}\,J(\bm{k}),
\end{equation}
{and}
\begin{equation}
	\mathcal{I}^\pm:=\mathcal{I}_\text{AA}\pm\mathcal{I}_\text{BB},
\end{equation}
and these integrals~(\ref{eq:IJtotal}) suffice to characterize~$\rho_{\text AB}$ fully.

{Straightforward} but lengthy algebra leads to
\begin{align}
	\rho_1^{(2)}
		=&\int\text{d}^3 \bm{k}\int\text{d}^3\bm k'\Big(I_{\text{A}}(\bm k){
			I_{\text{A}}(\bm k')}^*\proj{1(\bm{k})eg}{1(\bm{k}')eg}
				\nonumber\\&+
			I_{\text{B}}(\bm k){I_{\text{B}}(\bm k')}^*\proj{1(\bm{k})ge}{1(\bm{k}')ge}
				\nonumber\\
&+I_{\text{A}}(\bm k){I_{\text{B}}(\bm k')}^*\proj{1(\bm{k})eg}{1(\bm{k}')ge}\nonumber\\
&+I_{\text{B}}(\bm k){I_{\text{A}}(\bm k')}^*\proj{1(\bm{k})ge}{1(\bm{k}')eg}\Big)
\end{align}
with double excitations
{achieved} via counter-rotating term~$\hat{\sigma}^+\hat{a}^\dagger(\bm{k})$,
and higher states are not excited to this perturbation order.
A similar calculation {yields} $\rho_2^{(2)}$.
To compute {$\rho_\text{AB}$},
we trace out the field {to obtain}
\begin{align}
	\operatorname{tr}_\text{f}\rho_1^{(2)}
		=&\nonumber\int\text{d}^3 \bm k \,\Big[\big|I_{\text{A}}(\bm k)\big|^2 \proj{eg}{eg}+
			\big|I_{\text{B}}(\bm k)\big|^2\proj{ge}{ge}
				\nonumber\\&+
		\big(I_{\text{A}}(\bm k){I^*_{\text{B}}(\bm k')}\proj{eg}{ge}+\text{H.c.}\big)\Big],
				\nonumber\\
	\operatorname{tr}_\text{f}\rho_2^{(2)}
		=&-\int\text{d}^3 \bm k \,\Big[2\operatorname{Re}C(\bm k)\proj{gg}{gg}
				\nonumber\\&+
		J(\bm k)\proj{ee}{gg}
		+J^*(\bm k)\proj{gg}{ee}\Big],
\end{align}
and
$\int\text{d}^3 \bm k \,\operatorname{Re}C(\bm k)$
conveniently does not need to be calculated as it can be obtained from the other terms as each perturbative correction  is traceless order{-}by{-}order~\cite{Comm1}.

In the basis
$\left\{ \ket{gg},\ket{ge},\ket{eg},\ket{ee}\right\}$,
\begin{equation}
\label{eq:rhoAB}
	\rho_\text{AB}
		=\begin{pmatrix}
		1-\mathcal{I}^+ & 0 & 0 & -\mathcal{J}^*\\
		0 &\mathcal{I}_\text{BB} & \mathcal{I}_\text{AB}  &0\\
		0 & \mathcal{I}^*_\text{AB}  &\mathcal{I}_\text{AA} &0\\
		-\mathcal{J} &0 & 0 &0
		\end{pmatrix}+O(\lambda^4).
\end{equation}
In  contrast to one of the original entanglement-harvesting proposals
by Reznik et al.~\cite{RRS05},
one diagonal term vanishes here because exciting both detectors requires $O(\lambda^4)$ terms. If we were to include this term,
we would also have to include all the additional $O(\lambda^4)$ terms and not only the double-excitation one.
(Notice, however, that as Reznik et al.~\cite{RRS05} calculate entanglement to second order, all
their entanglement results remain correct in spite of this small inconsistency in the Dyson expansion.)

\section{Bipartite entanglement}
\label{sec:bipartite}

For a {two-qubit} system,
negativity~$\mathcal N$ conveniently estimates bipartite entanglement~\cite{Negat}. 
For transpose operator~$\textsf{T}$,
application of the partial transpose operation to a general density matrix yields
\begin{equation}
	\mathds{1}\otimes\textsf{T}:
	\sum_{ijkl}p^{ij}_{kl}\ket{i}\bra{j}\otimes\ket{k}\bra{l}
	\mapsto\sum_{ijkl}p^{ij}_{kl}\ket{i}\bra{j}\otimes\ket{l}\bra{k}
\end{equation}
whose negative eigenvalues sum to~$\mathcal N$.
From~(\ref{eq:rhoAB}),
\begin{equation}
\label{eq:PTrhoAB}
	\left(\mathds{1}\otimes\textsf{T}\right)\rho_{\text AB}
		\approx\begin{pmatrix}
			1-\mathcal{I}^+& 0 & 0 & \mathcal{I}_\text{AB}\\
			0 &\mathcal{I}_\text{BB} &  -\mathcal{J}^*  &0\\
			0 & -\mathcal{J}  &\mathcal{I}_\text{AA} &0\\
 			\mathcal{I}^*_\text{AB}   &0 & 0 &0
		\end{pmatrix}
		+O(\lambda^4)
\end{equation}
with only one negative eigenvalue at second order:
\begin{equation}
\label{eq:negatgen}
	\mathcal{N}
		=-\frac{1}{2}\left[\mathcal{I}^+
			-\sqrt{(\mathcal{I}^-)^2
					+4\left|\mathcal{J}\right|^2 }\right]+O(\lambda^4).
\end{equation}
For identical detectors switched on for the same amount of time ($\mathcal{I}^-=0$), 
negativity~(\ref{eq:negatgen}) simplifies to
\begin{equation}
\label{simpli}
	\mathcal{N}=\left|\mathcal{J}\right|-\mathcal{I}.
\end{equation}
As pointed out by Reznik et al.~\cite{RRS05},
intuitively the nonlocal term~$\mathcal{J}\propto\lambda_A\lambda_B$
must exceed local noise
$\mathcal{I}^+\propto\lambda^2$
for entanglement to arise.

We consider the case of sudden switching;
concomitant ultraviolet divergence is offset by spatial smearing~\cite{Langlois:2005nf,EMM2015}.
Switching is described by the rectangular function
\begin{equation}
	\sqcap(t)=\left\{\begin{matrix}
		1,&\text{if }|t|<1/2,\\
		1/2,&\text{if }|t|=1/2,\\
		0,&\text{if }|t|>1/2
		\end{matrix}\right.
\end{equation}
with Fourier transform
\begin{equation}
	\int_{-\infty}^\infty\text{d}t\sqcap(t)\text{e}^{-\text{i}\omega t}
		=\operatorname{sinc}\left(\frac{\omega}{2}\right).
\end{equation}
The detector is switched on at~$T$ and off at~$T'$ so
\begin{equation}
	\chi(t)
		=\sqcap
			\left(
				\frac{t-T}{2T^-}-\frac{1}{2}
			\right)
\end{equation}
for
\begin{equation}
	T^\pm:=\frac{T'\pm T}{2}.
\end{equation}

As
\begin{align}
	\frac{\text{e}^{\mathrm{i}\omega T'}-\text{e}^{\mathrm{i}\omega T}}{\omega}
		=&\frac{1}{\omega}\left.\text{e}^{\mathrm{i}\omega t}\right|_T^{T'}
			\nonumber\\
		=&2\mathrm{i}T^-\text{e}^{\mathrm{i}\omega T^+}
			\operatorname{sinc}\omega T^-,
\end{align}
we obtain 
\begin{align}
	I_{\nu}(\bm k)
		=&2\lambda_\nu T_\nu^-\text{e}^{\mathrm{i}\omega_\nu^+(\bm{k})T^+_\nu}
			\frac{\text{e}^{\mathrm{i} \bm k \cdot \bm x_\nu}}{\sqrt{2\omega}}
					\nonumber\\&\times
			\operatorname{sinc}(\omega_\nu^+(\bm{k}) T_\nu^-)\tilde F(\bm k)
\end{align}
so
\begin{equation}
\label{eq:Inunu}
	\mathcal{I}_{\nu\nu}
		= \int\text{d}^3 \bm k\,\frac{\lambda_\nu^2}{2\omega}\sin^2\left[\frac12\omega_\nu^+(\bm{k})
			 T_\nu^-\right]
		\left(\frac{2\tilde F(\bm k)}{\omega_\nu^+(\bm{k})}\right)^2.
\end{equation}
Making use of the compact support of~$\chi_\nu(t)$ over $[T_{\nu},T'_{\nu}]$,
we can write
\begin{equation}
\label{eq:jota}
	J(\bm k)
		=\lambda_\text{A}\lambda_\text{B}\text{e}^{\mathrm{i} \bm k\cdot\bm{r}}
			\left(\tilde J_\text{AB}(\bm k)+\tilde J_\text{BA}(\bm k)\right)
			\frac{\tilde F(\bm k)^2}{2\omega},
\end{equation}
where
\begin{equation}
	\tilde J_{\nu\nu'}(\bm k)
		:=\int_{T_\nu}^{T'_\nu}
			\text{d}t\int_{-\infty}^{t}\text{d}t'\,\chi_{\nu'}(t')
			\text{e}^{\mathrm{i}(\omega_{\nu'}^+(\bm{k})t'-\omega_\nu^-(\bm{k})t)}
\end{equation}
with $\nu,\nu'\in\{\text{A},\text{B}\}$.

{The} three distinct timing regimes are
\begin{enumerate}[(i)]
\item	$T'_\nu<T_{\nu'}$ with $\tilde J_{\nu\nu'}(\bm k)\equiv 0$;
\item$T'_{\nu'}<T_\nu$ for which
\begin{align}\label{eq:jota2}
	\tilde J_{\nu\nu'}(\bm k)
		=&-4 T_\nu^- T_{\nu'}^-
			\text{e}^{\mathrm{i}\left(\omega_{\nu'}^+(\bm k) T_{\nu'}^+
				-\omega_\nu^- T_\nu^+(\bm k)\right)}
					\nonumber\\&\times
				\operatorname{sinc}\omega_\nu^-(\bm k) T_\nu^-
				\operatorname{sinc}\omega_{\nu'}^+(\bm k) T_{\nu'}^+;
\end{align}
and
\item	the detector is on for overlapping times
for which we assume without loss of generality that
$T_{\nu'}<T_\nu$ and $T'_{\nu'}<T'_\nu$.
\end{enumerate}
As $\tilde J_{\nu\nu'}(\bm k)\neq0$ only if a detector is on,
we consider only the no-overlap time domain
\begin{equation}
	\mathcal{T}_0:=[T_{\nu'},T_\nu)\cup (T'_{\nu'},T'_\nu]
\end{equation}
or the overlapping interval
\begin{equation}
	\mathcal{T}_1:=[T_\nu,T'_{\nu'}].
\end{equation}
Thus,
\begin{equation}
	\tilde J_{\nu\nu'}=\tilde K_{\nu\nu'}-\tilde X_{\nu\nu'}
\end{equation}
with
\begin{align}
\label{eq:Kintegral}
	\tilde K_{\nu\nu'}(\bm k)
		=&	\frac{\left.\text{e}^{-\mathrm{i}\omega_\nu^-(\bm{k})t}\right|_{T_\nu}^{T'_\nu}}
				{\omega_\nu^-(\bm{k})}
			\frac{\left.\text{e}^{\mathrm{i}\omega_{\nu'}^+(\bm{k})t}\right|_{T_{\nu'}}^{T_\nu}}
				{\omega_{\nu'}^+(\bm{k})}
						\nonumber\\
		&+	\frac{\left.\text{e}^{-\mathrm{i}\omega_\nu^-(\bm{k})t}\right|_{T'_{\nu'}}^{T'_\nu}}
				{\omega_\nu^-(\bm{k})}
			\frac{\left.\text{e}^{\mathrm{i}\omega_{\nu'}^+(\bm{k})t}\right|_{T_{\nu}}^{T'_{\nu'}}}
				{\omega_{\nu'}^+(\bm{k})}
\end{align}
over~$\mathcal{T}_0$ and
\begin{align}
\label{eq:Xintegral}
	\tilde X_{\nu\nu'}(\bm k)
		=&	\frac{\text{e}^{\mathrm{i}
			\left(\omega_{\nu'}^+(\bm{k})T_\nu-\omega_\nu^-(\bm{k})T'_{\nu'}\right)}}
			{\omega_\nu^-(\bm{k})\omega_{\nu'}^+(\bm{k})}
				\nonumber\\&
			-\frac{\omega_{\nu'}^+(\bm{k})
			\text{e}^{\mathrm{i}T_\nu\Omega^+_{\nu\nu'}}-\omega_\nu^-(\bm{k})
			\text{e}^{\mathrm{i}T'_{\nu'}\Omega^+_{\nu\nu'}}}
			{\omega_\nu^-(\bm{k})\omega_{\nu'}^+(\bm{k})\Omega^+_{\nu\nu'}}
\end{align}
over~$\mathcal{T}_1$
for
\begin{equation}
	\Omega^+_{\nu\nu'}:=\Omega_{\nu} +\Omega_{\nu'}.
\end{equation}

We specialize to Gaussian spatial smearing of the detectors,
\begin{equation}
	F(\bm x)
		=\frac{\exp\left(-|\bm x|^2/\sigma^2\right)}{\left(\sqrt{\pi}\sigma\right)^3},
\end{equation}
which implies negligible overlap between smeared detectors
for reasonable separations $r\gg\sigma$.
{For} non-overlapping switching functions{,}
Eqs.~(\ref{eq:Inunu})-(\ref{eq:jota2}) and \eqref{eq:IJtotal} {yield}
\begin{align}
\label{eqs:IJGaussian}
	\mathcal{I}_{\nu\nu}
		=&\frac{\lambda_\nu^2}{\pi^2} \int_0^\infty\text{d}\omega
			\frac{\omega\text{e}^{-\frac12\omega^2\sigma^2}}{\omega_\nu^{+2}}
		\sin^2\left(\frac{\omega_\nu^+}{2}T^-_\nu\right),\\
	\left|\mathcal{J}\right|
		=&\frac{\lambda_\text{A}\lambda_\text{B}}{4\pi^2 r}
		\left|\int_0^\infty\text{d}\omega\sin\omega r
			\text{e}^{-\frac12\omega^2\sigma^2}
		\frac{\left.\text{e}^{-\mathrm{i}\omega_{\text B}^-t}\right|_{T_\text{B}}^{T'_\text{B}}}
			{\omega_{\text B}^-}
		\frac{\left.\text{e}^{\mathrm{i}\omega_{\text A}^+t}\right|_{T_\text{A}}^{T'_\text{A}}}
			{\omega_{\text A}^+}
		\right|	\nonumber
\end{align}
with important quantities
\begin{equation}
	\omega\sigma=|\bm{k}|\sigma
\end{equation}
being the wavenumber-spread product and 
$\omega r$ being the normalized separation in terms of wavenumber.
As~$\mathcal{J}$ is asymptotically proportional to $r^{-1}$,
entanglement,
which involves competition between $r$-independent~$\mathcal{I}_{\nu\nu}$
vs~$\mathcal{J}$,
decreases as~$r$ increases.

\begin{figure}
	(a)\includegraphics[width=.80\columnwidth]{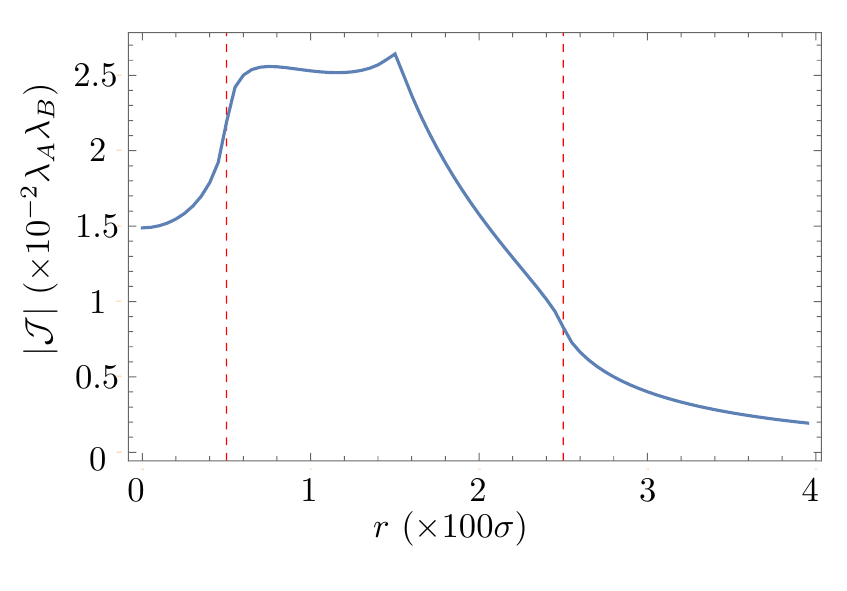}\\
	(b)\includegraphics[width=0.8\columnwidth]{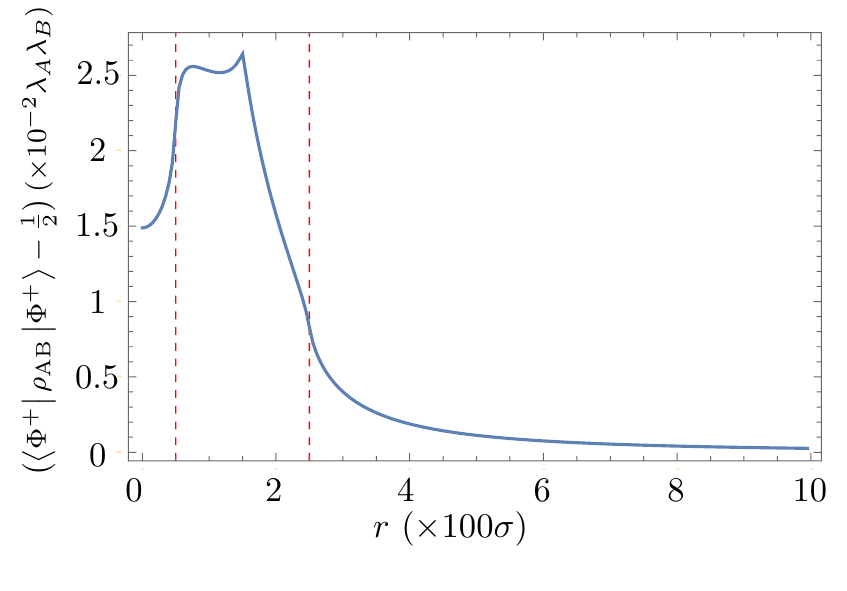}\\
	\caption{%
	(Color online)
	(a) Correlations term~$|\mathcal J|$ 
	and (b)~projection of~$\rho_\text{AB}$ onto~$\ket{\Phi^+}$
	vs distance~$r$ with $\Omega=1$
	for~ detectors with
	{$2T^-_\text{A,B}=100\sigma$, $T_\text{B}-T'_\text{A}=50\sigma$}, and $\sigma=0.001$,
	Vertical (red) dashed lines indicate the start and end of lightlike contact 
	for duration~$10\sigma$. Both peak on light-contact showing that the maximum of correlation between the atoms occurs at light-contact. Notice the leakage outside of the light-cone.%
	}
	\label{fig:harvesting}
\end{figure}

Although for simplicity and clarity we have chosen for our study
to employ a sudden switching function,
notice that the harvested entanglement in the case of a sudden switching is noticeably smaller than in the case of Gaussian switching~\cite{Pozas2015}. Note, however, that for any kind of switching, the fundamental feature which leads to entanglement harvesting (at leading order) is that the non-local term $\left|\mathcal{J}\right|$ dominates over the local noise $\mathcal{I}$ in
Eq.~\eqref{simpli}. 

\section{Sensitivity to space-time Positioning}

As is well understood (See, e.g.~\cite{Pozas2015}), harvested entanglement is evidently maximized for the two detectors being lightlike-separated
but can be nonzero even for detectors which remain spacelike separated. One can clearly see this by looking at the non-local term $\left|\mathcal{J}\right|$, which peaks if the two detectors are on for a long time while light-connected
but falls with increased~$r$.
This decline with respect to~$r$ applies to spacelike entanglement harvesting
for which detectors do not exchange real photons
and the correlation term $\left|\mathcal{J}\right|$ (and, therefore, the harvested entanglement) deteriorates due to {decay of} vacuum correlations with~$r$.

Additionally, for any entanglement in ~$\rho_\text{AB}$ to be useful,
Alice and Bob need an accurate description
of the mutually shared {(mixed)} state~$\rho_\text{AB}$,
which can be expressed in the Bell basis
\begin{align}
	\ket{\Phi^\pm}=&(1,0,0,\pm 1)^\textsf{T}/\sqrt{2},\nonumber\\
	\ket{\Psi^\pm}=&(0,1, \pm 1,0)^\textsf{T}/\sqrt{2}.
\end{align}
{The state's Bell-state fraction for the final state of the detectors is}
\begin{align}
	\bra{\Phi^\pm}\rho_\text{AB}\ket{\Phi^\pm}
		=&\frac12-\frac12\left[\mathcal{I}^-\mp2\text{Re}\left(\mathcal{J}\right)\right],
				\nonumber\\
	\bra{\Psi^\pm}\rho_\text{AB}\ket{\Psi^\pm}
		=&\frac12\left[\mathcal{I}^+\pm2\text{Re}\left(\mathcal{I}_\text{AB}\right)\right].
\end{align}
In Fig.~\ref{fig:harvesting} we show both the behaviour of the correlation term $\left|\mathcal{J}\right|$ and the projection of the shared correlated state onto $\ket{\Phi^+}$.
Evidently viable  entanglement harvesting is
maximized by the inter-detector light-crossing time being order unity expressed in terms of~$r$
in natural units.

{To exploit harvested} entanglement,
Alice and Bob {require an accurate description of the $r$-dependent} shared density matrix~$\rho_\text{AB}$,
but~$r$ necessarily has nonzero uncertainty~$\delta$.
We treat the joint state, subject to a mean separation~$r_0$
with uncertainty of~$\delta$,
as a smeared state
\begin{equation}
	\bar{\rho}_\text{AB}(\delta)=\int\!\text{d}r\operatorname{Pr}(r) \rho_\text{AB}(r),\;
	\operatorname{Pr}(r)=\frac{\text{e}^{-(r-r_0)^2/\delta^2}}{\delta\sqrt{\pi}},
\end{equation}
with the Gaussian distribution~Pr$(r)$ {formally} permitting an unphysical negative distance{,
which} is not problematic for reasonable separations.

Importantly, in the matrix elements of $\bar{\rho}_\text{AB}(\delta)$,
the only deleterious effect on~$\mathcal N$
arises in the term~$\mathcal{J}$
{as} the terms $\mathcal{I}^\pm$ {are} $r${-independent}
as expected from their local nature.
This difference allows us to focus the study of how synchronization affects the ability of Alice and Bob to harvest entanglement exclusively on the 
behaviour of the correlation term $\left|\mathcal{J}\right|$.

For the simpler non-overlapping case of the switching functions,
we can calculate a closed expression for $\left|\mathcal{J}\right|$ modulo the integral over field frequencies using the fact that
\begin{equation}
\label{eq:integralidentity}
\int_{-\infty}^\infty \frac{\text{d} r}{r}\frac{\sin(|\bm k| r)}{\text{e}^{\frac{(r-r_0)^2}{\delta^2}}}
	= \pi  \text{e}^{-\left(\frac{r_0}{\delta}\right)^2} \text{Im}\left[\text{Erfi}\left(\frac{r_0}{\delta }+\frac{\mathrm{i} \delta  |\bm k|}{2}\right)\right]
\end{equation}
where $\text{Erfi}(z)=-\mathrm{i}\text{Erf}(\mathrm{i} z)$ is the imaginary error function.
Taking expression~(\ref{eq:integralidentity}) into account,
the effect of a relative-positioning error~$\delta$ on the correlation term $\left|\mathcal{J}\right|$ is
\begin{align}
\label{eq:hopefullyfinal}
	\left|\mathcal{J}(\delta)\right|
		=&\frac{\lambda_\text{A}\lambda_\text{B}}{4\delta \pi^{\frac32}}
			\text{e}^{-\left(r_0/\delta\right)^2}
			\Bigg|\int_0^\infty\text{d}\omega\operatorname{Im}
			\left[\text{Erfi}\left(\frac{r_0}{\delta}+\frac{\mathrm{i} \delta\omega}{2}\right)\right]
					\nonumber\\&\times
		\text{e}^{-\frac12\omega^2\sigma^2}
		\frac{\left.\text{e}^{-\mathrm{i}\omega_{\text B}^-t}\right|_{T_\text{B}}^{T'_\text{B}}}
		{\omega_{\text B}^-}
		\frac{\left.\text{e}^{\mathrm{i}\omega_{\text A}^+t}\right|_{T_\text{A}}^{T'_\text{A}}}
		{\omega_{\text A}^+}
		\Bigg|,
\end{align}

We see from Eq.~(\ref{eq:hopefullyfinal}), that,
if the uncertainty of the inter-detector separation is fixed,
$\left|\mathcal{J}\right|$ decreases with decreasing spatial separation~$r_0$ between the atoms.
Specifically, the Gaussian term in Eq.~(\ref{eq:hopefullyfinal}) can dramatically reduce entanglement-harvesting capability for long distances.
On the other hand, if~$\delta$ increases,
we see that in the limit that~$\delta$ is larger than all the other length scales in the problem ($\delta\rightarrow\infty$),
both the Gaussian factor and the imaginary part of the Erfi function in Eq.~(\ref{eq:hopefullyfinal})  asymptotically approach unity, and then $\left|\mathcal{J}\right|\sim\delta^{-1}$. This shows that entanglement harvesting is impacted by spatial uncertainty in the positioning of the detector, and the decay is just of a polynomial kind.

If the two detectors held by Alice and Bob
are fixed precisely,
within precision~$\sigma$ for each detector,
prior to commencing the entanglement-harvesting protocol,
and~Alice and Bob are informed of the value~$r_0\gg\sigma$ to a high degree of precision,
then the effect of nonzero~$\delta$ can be ignored.
If, on the other hand,
Alice and Bob need to determine their separation as the first stage of the protocol,
which is a reasonable requirement for harvesting entanglement from the cosmological vacuum,
then establishing separation to precision within~$O(\sigma)$ independent of~$r$
is daunting.

We now analyze the resultant uncertainty due to fundamental imprecision
in synchronizing their respective reference frames.
Due to the covariant nature of the light-matter interaction, the impacts on the atom-field dynamics of a distance uncertainty and of a relative time uncertainty are the same
(as one of them becomes the other under an appropriate reference frame change).
This rule is true if the time uncertainty is insufficiently small to resolve the additional time scale in the problem that is not related,
in a Lorentz-covariant way, to a physical length scale (namely, $\Omega^{-1}$).
The reason the scale $\Omega^{-1}$ is different from the others  is that the internal dynamics of the atoms is considered (to a very good approximation) to be non-relativistic in the Unruh-Dewitt model (and any other usual models of light-matter interaction).

Although a lengthy task,
numerically checking verifies that,
for time uncertainties exceeding~$\Omega^{-1}$,
the uncertainties in space and time impact on entanglement harvesting 
in the same way as intimated by the above reasoning.
Thus, we focus our detailed analysis on the impact of a spatial uncertainty~$\delta$  on harvesting, which we  also call `synchronization error' for uncertainties much larger than the timescale $\Omega^{-1}$. As a last comment,  the scale of $\Omega^{-1}$ for the Hydrogen atom's first transition (of wavelength $\Lambda\sim 100$ nm) is $\Omega^{-1}=c^{-1}\Lambda\sim 10^{-16}$ s. The assumption that the main sources of uncertainty are going to be larger than this timescale is,
therefore, a good approximation, and henceforth we consider that imprecise determination of~$r$ and synchronization limitations are alike.

%



\begin{figure}
	\includegraphics[width=\columnwidth]{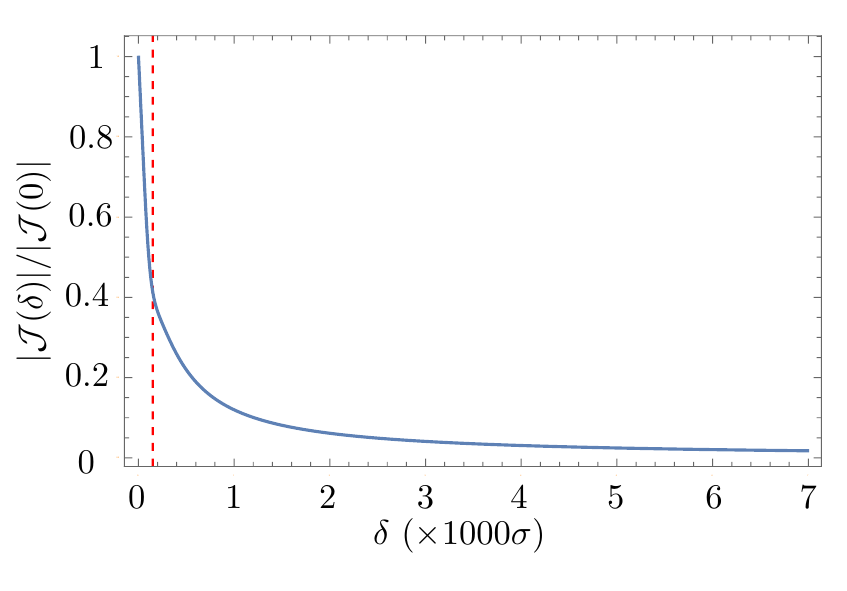}
	\caption{%
		(Color online)
		Correlation term~$|\mathcal J|$ (responsible for entanglement harvesting) vs separation imprecision~$\delta$
		for $r_0=150\sigma$, 
		$2T^-_\text{A,B}=100\sigma$, $T_\text{B}-T'_\text{A}=50\sigma$ and $\sigma=0.001$
		evaluated at optimal separation and switching time according to Fig.~\ref{fig:harvesting}(a).
		The vertical dashed (red) line represents the distance between the two detectors.
		Although the physically relevant region covers only the left of the red dashed line (uncertainties in positioning below the separation of the atoms, $\delta<r$) we plot the large uncertainty regime to show the polynomial decay of $|\mathcal J|$  with the imprecision~$\delta$.%
	}
	\label{fig:uncert}
\end{figure}

Alice's and Bob's harvestable entanglement from the field vacuum 
 is robust under imprecision
as exemplified in Fig.~\ref{fig:uncert},
where we plot the relative decrease in the correlation term as the imprecision in the determination of $r$ is increased. In particular we plot the ratio
\begin{equation}
	R(\delta)=\frac{|J(\delta)|}{|J(0)|}
\end{equation}
evaluated in the vicinity of the peak in Fig.~\ref{fig:harvesting}. We see that the correlation term $|\mathcal{J}(\delta)|$ decays 60\% when the uncertainty in the determination of relative distance is the separation of the detectors.
Notice that in most experiments the uncertainty in the separation between the two atoms would be smaller than the distance between the atoms $r$, and typically larger than the atomic size $\sigma$, that is $r>\delta\gg\sigma$. This means that the most physically relevant region of the plot is to the left of the red dashed line.  We plot the behaviour of the correlation term $|\mathcal{J}(\delta)|$ for uncertainties larger than the separation between the detectors to illustrate the slow (polynomial) decay of the ability of the setup to harvest entanglement.

When the uncertainty~$\delta$ is larger than the distance between the detectors, one may think that  Fig.~\ref{fig:harvesting} may contain non-physical situations where the two detectors are overlapping. However, as $r\ll\sigma$, if we think of the uncertainty emerging from  an ensemble of experiments where the positions of the centre of mass of the two detectors in every experiment of the ensemble are randomly set along the Gaussian-distributed length~$\delta$,
the probability of the two detectors landing in a region where they would have a non-negligible overlap (of the order of a few $\sigma$) is exponentially suppressed. Therefore, the effect of these possible superpositions of the two detectors on features of   the plot in Fig.~\ref{fig:harvesting}  is  negligible.

It is known that, even in the cases where entanglement harvesting is possible for a broad set of physical parameters (See, e.g.,~\cite{Pozas2015}), the correlation term~$|\mathcal{J}|$ never becomes much larger than the noise term~$\mathcal{I}$.
Instead, in all scenarios for which entanglement harvesting is possible, the correlation term becomes larger but of the same order of magnitude than the noise term. 

For entanglement harvesting, $|\mathcal{J}|>\mathcal{I}$ is crucial,
and damping by~$50\%$ of the correlation term already suffices to prevent entanglement harvesting from happening.
Thus,
according to our results,
if the uncertainty in the determination of the spacetime synchronization is of the order of the separation between the detectors, entanglement harvesting is impeded.

 As we can see in \eqref{eq:hopefullyfinal}, the decay of the term  $|\mathcal{J}|$ is dictated by the ratio of the spatial separation to the distance uncertainty ($r_0/\delta$) and not the smearing length scale of the detectors $\sigma$. This means that, even though in Fig.~\ref{fig:uncert} we show  regimes where  $\delta\gg\sigma$, the decay in the non-local term $|\mathcal{J}|$ is entirely due to the increase of uncertainty~$\delta$ and not to any kind of overlap between the detector smearings or any other effect related to the size of the detectors relative to the magnitude of their spatial separation.
 
\section{Conclusions}

We have reviewed that  entanglement harvesting is the result of the the non-local terms $|\mathcal{J}|$ dominating over the local noise terms $\mathcal{I}$ in the interaction of a system of two detectors with the vacuum state of a quantum field. We have analyzed how imprecisions on the space-time synchronization of the two detectors (their relative distance and their ability to synchronize their clocks) affects both the local noise and the non-local terms.

We have shown that while, of course, the local noise terms are unaffected by imprecision in the synchronization of the detectors, the non-local term gets exponentially damped in the square of the relative uncertainty in the space-time positioning. This means that in scenarios where entanglement harvesting is possible, the harvesting of entanglement between two particle detectors is robust under relative imprecisions in the spatial and temporal positioning if they are below the scale of their spatial separation and the light crossing time between them.
In contrast, the ability to harvest entanglement deteriorates rapidly when the precision  in the space-time positioning becomes non-negligible with respect to the separation between detectors.

If  the precision in spacetime positioning is comparable to the size of the detectors $~\sigma\ll r_0$,
as could be the case of atomic-scale detectors in a well controlled table-top experiment, our results hint that vacuum entanglement harvesting would be robust under distance-time measurement imprecisions. On astronomical scales, however,
in situ measurements of distance are generally not possible so distances are instead inferred
from various measured quantities such as the product of redshift and Hubble's constant~\cite{Hub29}
where even establishing the value of the constant is challenging~\cite{FMG+01}.
This limitation indicates that our results,
when extrapolated to a cosmological scenario where the field is conformally coupled to curvature,
could be particularly relevant
when considering entanglement harvesting in cosmological contexts~\cite{VerSteeg2009,Blasco2015}.

\acknowledgments
BCS acknowledges financial support from AITF, NSERC and China's 1000 Talent Plan. E. M-M.  acknowledges financial support through NSERC Discovery programme. E. M-M would like to thank A. Pozas for helpful discussions. E. M-M thanks A. Blasco, M. Mart\'in-Benito and L. J. Garay for their contributions and suggestions regarding the final presentation of this article.
\bibliography{biblio}

\begin{thebibliography}{31}%
\makeatletter
\providecommand \@ifxundefined [1]{%
 \@ifx{#1\undefined}
}%
\providecommand \@ifnum [1]{%
 \ifnum #1\expandafter \@firstoftwo
 \else \expandafter \@secondoftwo
 \fi
}%
\providecommand \@ifx [1]{%
 \ifx #1\expandafter \@firstoftwo
 \else \expandafter \@secondoftwo
 \fi
}%
\providecommand \natexlab [1]{#1}%
\providecommand \enquote  [1]{``#1''}%
\providecommand \bibnamefont  [1]{#1}%
\providecommand \bibfnamefont [1]{#1}%
\providecommand \citenamefont [1]{#1}%
\providecommand \href@noop [0]{\@secondoftwo}%
\providecommand \href [0]{\begingroup \@sanitize@url \@href}%
\providecommand \@href[1]{\@@startlink{#1}\@@href}%
\providecommand \@@href[1]{\endgroup#1\@@endlink}%
\providecommand \@sanitize@url [0]{\catcode `\\12\catcode `\$12\catcode
  `\&12\catcode `\#12\catcode `\^12\catcode `\_12\catcode `\%12\relax}%
\providecommand \@@startlink[1]{}%
\providecommand \@@endlink[0]{}%
\providecommand \url  [0]{\begingroup\@sanitize@url \@url }%
\providecommand \@url [1]{\endgroup\@href {#1}{\urlprefix }}%
\providecommand \urlprefix  [0]{URL }%
\providecommand \Eprint [0]{\href }%
\providecommand \doibase [0]{http://dx.doi.org/}%
\providecommand \selectlanguage [0]{\@gobble}%
\providecommand \bibinfo  [0]{\@secondoftwo}%
\providecommand \bibfield  [0]{\@secondoftwo}%
\providecommand \translation [1]{[#1]}%
\providecommand \BibitemOpen [0]{}%
\providecommand \bibitemStop [0]{}%
\providecommand \bibitemNoStop [0]{.\EOS\space}%
\providecommand \EOS [0]{\spacefactor3000\relax}%
\providecommand \BibitemShut  [1]{\csname bibitem#1\endcsname}%
\let\auto@bib@innerbib\@empty
\bibitem [{\citenamefont {Unruh}(1976)}]{Unr76}%
  \BibitemOpen
  \bibfield  {author} {\bibinfo {author} {\bibfnamefont {W.~G.}\ \bibnamefont
  {Unruh}},\ }\href {\doibase 10.1103/PhysRevD.14.870} {\bibfield  {journal}
  {\bibinfo  {journal} {Phys. Rev. D}\ }\textbf {\bibinfo {volume} {14}},\
  \bibinfo {pages} {870} (\bibinfo {year} {1976})}\BibitemShut {NoStop}%
\bibitem [{\citenamefont {Valentini}(1991)}]{Valentini1991}%
  \BibitemOpen
  \bibfield  {author} {\bibinfo {author} {\bibfnamefont {A.}~\bibnamefont
  {Valentini}},\ }\href {\doibase
  http://dx.doi.org/10.1016/0375-9601(91)90952-5} {\bibfield  {journal}
  {\bibinfo  {journal} {Physics Letters A}\ }\textbf {\bibinfo {volume}
  {153}},\ \bibinfo {pages} {321 } (\bibinfo {year} {1991})}\BibitemShut
  {NoStop}%
\bibitem [{\citenamefont {Reznik}(2003)}]{Rez03}%
  \BibitemOpen
  \bibfield  {author} {\bibinfo {author} {\bibfnamefont {B.}~\bibnamefont
  {Reznik}},\ }\href@noop {} {\bibfield  {journal} {\bibinfo  {journal} {Found.
  Phys.}\ }\textbf {\bibinfo {volume} {33}},\ \bibinfo {pages} {167} (\bibinfo
  {year} {2003})}\BibitemShut {NoStop}%
\bibitem [{\citenamefont {Reznik}\ \emph {et~al.}(2005)\citenamefont {Reznik},
  \citenamefont {Retzker},\ and\ \citenamefont {Silman}}]{RRS05}%
  \BibitemOpen
  \bibfield  {author} {\bibinfo {author} {\bibfnamefont {B.}~\bibnamefont
  {Reznik}}, \bibinfo {author} {\bibfnamefont {A.}~\bibnamefont {Retzker}}, \
  and\ \bibinfo {author} {\bibfnamefont {J.}~\bibnamefont {Silman}},\ }\href
  {\doibase 10.1103/PhysRevA.71.042104} {\bibfield  {journal} {\bibinfo
  {journal} {Phys. Rev. A}\ }\textbf {\bibinfo {volume} {71}},\ \bibinfo
  {pages} {042104} (\bibinfo {year} {2005})}\BibitemShut {NoStop}%
\bibitem [{\citenamefont {Salton}\ \emph {et~al.}(2015)\citenamefont {Salton},
  \citenamefont {Mann},\ and\ \citenamefont {Menicucci}}]{Salton:2014jaa}%
  \BibitemOpen
  \bibfield  {author} {\bibinfo {author} {\bibfnamefont {G.}~\bibnamefont
  {Salton}}, \bibinfo {author} {\bibfnamefont {R.~B.}\ \bibnamefont {Mann}}, \
  and\ \bibinfo {author} {\bibfnamefont {N.~C.}\ \bibnamefont {Menicucci}},\
  }\href {\doibase 10.1088/1367-2630/17/3/035001} {\bibfield  {journal}
  {\bibinfo  {journal} {New J. Phys.}\ }\textbf {\bibinfo {volume} {17}},\
  \bibinfo {pages} {035001} (\bibinfo {year} {2015})}\BibitemShut {NoStop}%
\bibitem [{\citenamefont {Pozas-Kerstjens}\ and\ \citenamefont
  {Mart\'{\i}n-Mart\'{\i}nez}(2015)}]{Pozas2015}%
  \BibitemOpen
  \bibfield  {author} {\bibinfo {author} {\bibfnamefont {A.}~\bibnamefont
  {Pozas-Kerstjens}}\ and\ \bibinfo {author} {\bibfnamefont {E.}~\bibnamefont
  {Mart\'{\i}n-Mart\'{\i}nez}},\ }\href {\doibase 10.1103/PhysRevD.92.064042}
  {\bibfield  {journal} {\bibinfo  {journal} {Phys. Rev. D}\ }\textbf {\bibinfo
  {volume} {92}},\ \bibinfo {pages} {064042} (\bibinfo {year}
  {2015})}\BibitemShut {NoStop}%
\bibitem [{\citenamefont {Bennett}\ \emph {et~al.}(1993)\citenamefont
  {Bennett}, \citenamefont {Brassard}, \citenamefont {Cr\'epeau}, \citenamefont
  {Jozsa}, \citenamefont {Peres},\ and\ \citenamefont {Wootters}}]{BBC+93}%
  \BibitemOpen
  \bibfield  {author} {\bibinfo {author} {\bibfnamefont {C.}~\bibnamefont
  {Bennett}}, \bibinfo {author} {\bibfnamefont {G.}~\bibnamefont {Brassard}},
  \bibinfo {author} {\bibfnamefont {C.}~\bibnamefont {Cr\'epeau}}, \bibinfo
  {author} {\bibfnamefont {R.}~\bibnamefont {Jozsa}}, \bibinfo {author}
  {\bibfnamefont {A.}~\bibnamefont {Peres}}, \ and\ \bibinfo {author}
  {\bibfnamefont {W.}~\bibnamefont {Wootters}},\ }\href {\doibase
  10.1103/PhysRevLett.70.1895} {\bibfield  {journal} {\bibinfo  {journal}
  {Phys. Rev. Lett.}\ }\textbf {\bibinfo {volume} {70}},\ \bibinfo {pages}
  {1895} (\bibinfo {year} {1993})}\BibitemShut {NoStop}%
\bibitem [{\citenamefont {Bouwmeester}\ \emph {et~al.}(1997)\citenamefont
  {Bouwmeester}, \citenamefont {Pan}, \citenamefont {Mattle}, \citenamefont
  {Eibl}, \citenamefont {Weinfurter},\ and\ \citenamefont
  {Zeilinger}}]{BPM+97}%
  \BibitemOpen
  \bibfield  {author} {\bibinfo {author} {\bibfnamefont {D.}~\bibnamefont
  {Bouwmeester}}, \bibinfo {author} {\bibfnamefont {J.-W.}\ \bibnamefont
  {Pan}}, \bibinfo {author} {\bibfnamefont {K.}~\bibnamefont {Mattle}},
  \bibinfo {author} {\bibfnamefont {M.}~\bibnamefont {Eibl}}, \bibinfo {author}
  {\bibfnamefont {H.}~\bibnamefont {Weinfurter}}, \ and\ \bibinfo {author}
  {\bibfnamefont {A.}~\bibnamefont {Zeilinger}},\ }\href@noop {} {\bibfield
  {journal} {\bibinfo  {journal} {Nature}\ }\textbf {\bibinfo {volume} {390}},\
  \bibinfo {pages} {575} (\bibinfo {year} {1997})}\BibitemShut {NoStop}%
\bibitem [{\citenamefont {Bennett}\ and\ \citenamefont {Wiesner}(1992)}]{BW92}%
  \BibitemOpen
  \bibfield  {author} {\bibinfo {author} {\bibfnamefont {C.~H.}\ \bibnamefont
  {Bennett}}\ and\ \bibinfo {author} {\bibfnamefont {S.~J.}\ \bibnamefont
  {Wiesner}},\ }\href {\doibase 10.1103/PhysRevLett.69.2881} {\bibfield
  {journal} {\bibinfo  {journal} {Phys. Rev. Lett.}\ }\textbf {\bibinfo
  {volume} {69}},\ \bibinfo {pages} {2881} (\bibinfo {year}
  {1992})}\BibitemShut {NoStop}%
\bibitem [{\citenamefont {Mattle}\ \emph {et~al.}(1996)\citenamefont {Mattle},
  \citenamefont {Weinfurter}, \citenamefont {Kwiat},\ and\ \citenamefont
  {Zeilinger}}]{MWKZ96}%
  \BibitemOpen
  \bibfield  {author} {\bibinfo {author} {\bibfnamefont {K.}~\bibnamefont
  {Mattle}}, \bibinfo {author} {\bibfnamefont {H.}~\bibnamefont {Weinfurter}},
  \bibinfo {author} {\bibfnamefont {P.}~\bibnamefont {Kwiat}}, \ and\ \bibinfo
  {author} {\bibfnamefont {A.}~\bibnamefont {Zeilinger}},\ }\href {\doibase
  10.1103/PhysRevLett.76.4656} {\bibfield  {journal} {\bibinfo  {journal}
  {Phys. Rev. Lett.}\ }\textbf {\bibinfo {volume} {76}},\ \bibinfo {pages}
  {4656} (\bibinfo {year} {1996})}\BibitemShut {NoStop}%
\bibitem [{\citenamefont {Buhrman}\ \emph {et~al.}(2001)\citenamefont
  {Buhrman}, \citenamefont {Cleve}, \citenamefont {Watrous},\ and\
  \citenamefont {de~Wolf}}]{BCWdW01}%
  \BibitemOpen
  \bibfield  {author} {\bibinfo {author} {\bibfnamefont {H.}~\bibnamefont
  {Buhrman}}, \bibinfo {author} {\bibfnamefont {R.}~\bibnamefont {Cleve}},
  \bibinfo {author} {\bibfnamefont {J.}~\bibnamefont {Watrous}}, \ and\
  \bibinfo {author} {\bibfnamefont {R.}~\bibnamefont {de~Wolf}},\ }\href
  {\doibase 10.1103/PhysRevLett.87.167902} {\bibfield  {journal} {\bibinfo
  {journal} {Phys. Rev. Lett.}\ }\textbf {\bibinfo {volume} {87}},\ \bibinfo
  {pages} {167902} (\bibinfo {year} {2001})}\BibitemShut {NoStop}%
\bibitem [{\citenamefont {Horn}\ \emph {et~al.}(2005)\citenamefont {Horn},
  \citenamefont {Babichev}, \citenamefont {Marzlin}, \citenamefont {Lvovsky},\
  and\ \citenamefont {Sanders}}]{HBM+05}%
  \BibitemOpen
  \bibfield  {author} {\bibinfo {author} {\bibfnamefont {R.~T.}\ \bibnamefont
  {Horn}}, \bibinfo {author} {\bibfnamefont {S.~A.}\ \bibnamefont {Babichev}},
  \bibinfo {author} {\bibfnamefont {K.-P.}\ \bibnamefont {Marzlin}}, \bibinfo
  {author} {\bibfnamefont {A.~I.}\ \bibnamefont {Lvovsky}}, \ and\ \bibinfo
  {author} {\bibfnamefont {B.~C.}\ \bibnamefont {Sanders}},\ }\href {\doibase
  10.1103/PhysRevLett.95.150502} {\bibfield  {journal} {\bibinfo  {journal}
  {Phys. Rev. Lett.}\ }\textbf {\bibinfo {volume} {95}},\ \bibinfo {pages}
  {150502} (\bibinfo {year} {2005})}\BibitemShut {NoStop}%
\bibitem [{\citenamefont {Mart\'{i}n-Mart\'{i}nez}\ \emph
  {et~al.}(2013{\natexlab{a}})\citenamefont {Mart\'{i}n-Mart\'{i}nez},
  \citenamefont {Brown}, \citenamefont {Donnelly},\ and\ \citenamefont
  {Kempf}}]{MBDK13}%
  \BibitemOpen
  \bibfield  {author} {\bibinfo {author} {\bibfnamefont {E.}~\bibnamefont
  {Mart\'{i}n-Mart\'{i}nez}}, \bibinfo {author} {\bibfnamefont
  {E.}~\bibnamefont {Brown}}, \bibinfo {author} {\bibfnamefont
  {W.}~\bibnamefont {Donnelly}}, \ and\ \bibinfo {author} {\bibfnamefont
  {A.}~\bibnamefont {Kempf}},\ }\href {\doibase 10.1103/PhysRevA.88.052310}
  {\bibfield  {journal} {\bibinfo  {journal} {Phys. Rev. A}\ }\textbf {\bibinfo
  {volume} {88}},\ \bibinfo {pages} {052310} (\bibinfo {year}
  {2013}{\natexlab{a}})}\BibitemShut {NoStop}%
\bibitem [{\citenamefont {DeWitt}(1980)}]{DeW79}%
  \BibitemOpen
  \bibfield  {author} {\bibinfo {author} {\bibfnamefont {B.~S.}\ \bibnamefont
  {DeWitt}},\ }\enquote {\bibinfo {title} {General relativity: An einstein
  centenary survey},}\ \ (\bibinfo  {publisher} {Cambridge University Press},\
  \bibinfo {address} {Cambridge UK},\ \bibinfo {year} {1980})\ pp.\ \bibinfo
  {pages} {680--745}\BibitemShut {NoStop}%
\bibitem [{\citenamefont {Louko}\ and\ \citenamefont {Satz}(2006)}]{LS06}%
  \BibitemOpen
  \bibfield  {author} {\bibinfo {author} {\bibfnamefont {J.}~\bibnamefont
  {Louko}}\ and\ \bibinfo {author} {\bibfnamefont {A.}~\bibnamefont {Satz}},\
  }\href {\doibase 10.1088/0264-9381/23/22/015} {\bibfield  {journal} {\bibinfo
   {journal} {Class. Quant. Grav.}\ }\textbf {\bibinfo {volume} {23}},\
  \bibinfo {pages} {6321} (\bibinfo {year} {2006})}\BibitemShut {NoStop}%
\bibitem [{\citenamefont {Louko}\ and\ \citenamefont {Satz}(2008)}]{LS08}%
  \BibitemOpen
  \bibfield  {author} {\bibinfo {author} {\bibfnamefont {J.}~\bibnamefont
  {Louko}}\ and\ \bibinfo {author} {\bibfnamefont {A.}~\bibnamefont {Satz}},\
  }\href {\doibase 10.1088/0264-9381/25/5/055012} {\bibfield  {journal}
  {\bibinfo  {journal} {Class. Quant. Grav.}\ }\textbf {\bibinfo {volume}
  {25}},\ \bibinfo {pages} {055012} (\bibinfo {year} {2008})}\BibitemShut
  {NoStop}%
\bibitem [{\citenamefont {Mart\'{i}n-Mart\'{i}nez}\ \emph
  {et~al.}(2013{\natexlab{b}})\citenamefont {Mart\'{i}n-Mart\'{i}nez},
  \citenamefont {Montero},\ and\ \citenamefont {del Rey}}]{Wavepackets}%
  \BibitemOpen
  \bibfield  {author} {\bibinfo {author} {\bibfnamefont {E.}~\bibnamefont
  {Mart\'{i}n-Mart\'{i}nez}}, \bibinfo {author} {\bibfnamefont
  {M.}~\bibnamefont {Montero}}, \ and\ \bibinfo {author} {\bibfnamefont
  {M.}~\bibnamefont {del Rey}},\ }\href {\doibase 10.1103/PhysRevD.87.064038}
  {\bibfield  {journal} {\bibinfo  {journal} {Phys. Rev. D}\ }\textbf {\bibinfo
  {volume} {87}},\ \bibinfo {pages} {064038} (\bibinfo {year}
  {2013}{\natexlab{b}})}\BibitemShut {NoStop}%
\bibitem [{\citenamefont {Alhambra}\ \emph {et~al.}(2014)\citenamefont
  {Alhambra}, \citenamefont {Kempf},\ and\ \citenamefont
  {Mart\'in-Mart\'inez}}]{Alvaro}%
  \BibitemOpen
  \bibfield  {author} {\bibinfo {author} {\bibfnamefont {A.~M.}\ \bibnamefont
  {Alhambra}}, \bibinfo {author} {\bibfnamefont {A.}~\bibnamefont {Kempf}}, \
  and\ \bibinfo {author} {\bibfnamefont {E.}~\bibnamefont
  {Mart\'in-Mart\'inez}},\ }\href {\doibase 10.1103/PhysRevA.89.033835}
  {\bibfield  {journal} {\bibinfo  {journal} {Phys. Rev. A}\ }\textbf {\bibinfo
  {volume} {89}},\ \bibinfo {pages} {033835} (\bibinfo {year}
  {2014})}\BibitemShut {NoStop}%
\bibitem [{\citenamefont {Vidal}\ and\ \citenamefont {Werner}(2002)}]{Negat}%
  \BibitemOpen
  \bibfield  {author} {\bibinfo {author} {\bibfnamefont {G.}~\bibnamefont
  {Vidal}}\ and\ \bibinfo {author} {\bibfnamefont {R.~F.}\ \bibnamefont
  {Werner}},\ }\href@noop {} {\bibfield  {journal} {\bibinfo  {journal} {Phys.
  Rev. A}\ }\textbf {\bibinfo {volume} {65}},\ \bibinfo {pages} {032314}
  (\bibinfo {year} {2002})}\BibitemShut {NoStop}%
\bibitem [{\citenamefont {\ifmmode~\dot{Z}\else \.{Z}\fi{}yczkowski}\ \emph
  {et~al.}(1998)\citenamefont {\ifmmode~\dot{Z}\else \.{Z}\fi{}yczkowski},
  \citenamefont {Horodecki}, \citenamefont {Sanpera},\ and\ \citenamefont
  {Lewenstein}}]{ZHSL98}%
  \BibitemOpen
  \bibfield  {author} {\bibinfo {author} {\bibfnamefont {K.}~\bibnamefont
  {\ifmmode~\dot{Z}\else \.{Z}\fi{}yczkowski}}, \bibinfo {author}
  {\bibfnamefont {P.}~\bibnamefont {Horodecki}}, \bibinfo {author}
  {\bibfnamefont {A.}~\bibnamefont {Sanpera}}, \ and\ \bibinfo {author}
  {\bibfnamefont {M.}~\bibnamefont {Lewenstein}},\ }\href {\doibase
  10.1103/PhysRevA.58.883} {\bibfield  {journal} {\bibinfo  {journal} {Phys.
  Rev. A}\ }\textbf {\bibinfo {volume} {58}},\ \bibinfo {pages} {883} (\bibinfo
  {year} {1998})}\BibitemShut {NoStop}%
\bibitem [{\citenamefont {Plenio}(2005)}]{Ple05}%
  \BibitemOpen
  \bibfield  {author} {\bibinfo {author} {\bibfnamefont {M.}~\bibnamefont
  {Plenio}},\ }\href {\doibase 10.1103/PhysRevLett.95.090503} {\bibfield
  {journal} {\bibinfo  {journal} {Phys. Rev. Lett.}\ }\textbf {\bibinfo
  {volume} {95}},\ \bibinfo {pages} {090503} (\bibinfo {year}
  {2005})}\BibitemShut {NoStop}%
\bibitem [{\citenamefont {Jonsson}\ \emph {et~al.}(2015)\citenamefont
  {Jonsson}, \citenamefont {Mart\'{\i}n-Mart\'{\i}nez},\ and\ \citenamefont
  {Kempf}}]{Comm2}%
  \BibitemOpen
  \bibfield  {author} {\bibinfo {author} {\bibfnamefont {R.~H.}\ \bibnamefont
  {Jonsson}}, \bibinfo {author} {\bibfnamefont {E.}~\bibnamefont
  {Mart\'{\i}n-Mart\'{\i}nez}}, \ and\ \bibinfo {author} {\bibfnamefont
  {A.}~\bibnamefont {Kempf}},\ }\href {\doibase 10.1103/PhysRevLett.114.110505}
  {\bibfield  {journal} {\bibinfo  {journal} {Phys. Rev. Lett.}\ }\textbf
  {\bibinfo {volume} {114}},\ \bibinfo {pages} {110505} (\bibinfo {year}
  {2015})}\BibitemShut {NoStop}%
\bibitem [{\citenamefont {Blasco}\ \emph {et~al.}(2015)\citenamefont {Blasco},
  \citenamefont {Garay}, \citenamefont {Mart\'{\i}n-Benito},\ and\
  \citenamefont {Mart\'{\i}n-Mart\'{\i}nez}}]{Blasco2015}%
  \BibitemOpen
  \bibfield  {author} {\bibinfo {author} {\bibfnamefont {A.}~\bibnamefont
  {Blasco}}, \bibinfo {author} {\bibfnamefont {L.~J.}\ \bibnamefont {Garay}},
  \bibinfo {author} {\bibfnamefont {M.}~\bibnamefont {Mart\'{\i}n-Benito}}, \
  and\ \bibinfo {author} {\bibfnamefont {E.}~\bibnamefont
  {Mart\'{\i}n-Mart\'{\i}nez}},\ }\href {\doibase
  10.1103/PhysRevLett.114.141103} {\bibfield  {journal} {\bibinfo  {journal}
  {Phys. Rev. Lett.}\ }\textbf {\bibinfo {volume} {114}},\ \bibinfo {pages}
  {141103} (\bibinfo {year} {2015})}\BibitemShut {NoStop}%
\bibitem [{\citenamefont {Mart\'{\i}n-Mart\'{\i}nez}(2015)}]{EMM2015}%
  \BibitemOpen
  \bibfield  {author} {\bibinfo {author} {\bibfnamefont {E.}~\bibnamefont
  {Mart\'{\i}n-Mart\'{\i}nez}},\ }\href {\doibase 10.1103/PhysRevD.92.104019}
  {\bibfield  {journal} {\bibinfo  {journal} {Phys. Rev. D}\ }\textbf {\bibinfo
  {volume} {92}},\ \bibinfo {pages} {104019} (\bibinfo {year}
  {2015})}\BibitemShut {NoStop}%
\bibitem [{\citenamefont {Jonsson}(2015)}]{JonssonS}%
  \BibitemOpen
  \bibfield  {author} {\bibinfo {author} {\bibfnamefont {R.~H.}\ \bibnamefont
  {Jonsson}},\ }\href@noop {} {\enquote {\bibinfo {title} {Information travels
  in massless fields in 1+1 dimensions where energy cannot},}\ } (\bibinfo
  {year} {2015})\BibitemShut {NoStop}%
\bibitem [{\citenamefont {Blasco}\ \emph {et~al.}(2016)\citenamefont {Blasco},
  \citenamefont {Garay}, \citenamefont {Mart\'{\i}n-Benito},\ and\
  \citenamefont {Mart\'{\i}n-Mart\'{\i}nez}}]{Blasco2016}%
  \BibitemOpen
  \bibfield  {author} {\bibinfo {author} {\bibfnamefont {A.}~\bibnamefont
  {Blasco}}, \bibinfo {author} {\bibfnamefont {L.~J.}\ \bibnamefont {Garay}},
  \bibinfo {author} {\bibfnamefont {M.}~\bibnamefont {Mart\'{\i}n-Benito}}, \
  and\ \bibinfo {author} {\bibfnamefont {E.}~\bibnamefont
  {Mart\'{\i}n-Mart\'{\i}nez}},\ }\href {\doibase 10.1103/PhysRevD.93.024055}
  {\bibfield  {journal} {\bibinfo  {journal} {Phys. Rev. D}\ }\textbf {\bibinfo
  {volume} {93}},\ \bibinfo {pages} {024055} (\bibinfo {year}
  {2016})}\BibitemShut {NoStop}%
\bibitem [{\citenamefont {Jonsson}\ \emph {et~al.}(2014)\citenamefont
  {Jonsson}, \citenamefont {Mart\'{i}n-Mart\'{i}nez},\ and\ \citenamefont
  {Kempf}}]{Comm1}%
  \BibitemOpen
  \bibfield  {author} {\bibinfo {author} {\bibfnamefont {R.~H.}\ \bibnamefont
  {Jonsson}}, \bibinfo {author} {\bibfnamefont {E.}~\bibnamefont
  {Mart\'{i}n-Mart\'{i}nez}}, \ and\ \bibinfo {author} {\bibfnamefont
  {A.}~\bibnamefont {Kempf}},\ }\href {\doibase 10.1103/PhysRevA.89.022330}
  {\bibfield  {journal} {\bibinfo  {journal} {Phys. Rev. A}\ }\textbf {\bibinfo
  {volume} {89}},\ \bibinfo {pages} {022330} (\bibinfo {year}
  {2014})}\BibitemShut {NoStop}%
\bibitem [{\citenamefont {Langlois}(2006)}]{Langlois:2005nf}%
  \BibitemOpen
  \bibfield  {author} {\bibinfo {author} {\bibfnamefont {P.}~\bibnamefont
  {Langlois}},\ }\href {\doibase 10.1016/j.aop.2006.01.013} {\bibfield
  {journal} {\bibinfo  {journal} {Annals Phys.}\ }\textbf {\bibinfo {volume}
  {321}},\ \bibinfo {pages} {2027} (\bibinfo {year} {2006})}\BibitemShut
  {NoStop}%
\bibitem [{\citenamefont {Hubble}(1929)}]{Hub29}%
  \BibitemOpen
  \bibfield  {author} {\bibinfo {author} {\bibfnamefont {E.}~\bibnamefont
  {Hubble}},\ }\href {\doibase 10.1073/pnas.15.3.168} {\bibfield  {journal}
  {\bibinfo  {journal} {Proc. Natl. Acad. Sci. U.S.A.}\ }\textbf {\bibinfo
  {volume} {15}},\ \bibinfo {pages} {168} (\bibinfo {year} {1929})}\BibitemShut
  {NoStop}%
\bibitem [{\citenamefont {Freedman}\ \emph {et~al.}(2001)\citenamefont
  {Freedman}, \citenamefont {Madore}, \citenamefont {Gibson}, \citenamefont
  {Ferrarese}, \citenamefont {Kelson}, \citenamefont {Sakai}, \citenamefont
  {Mould}, \citenamefont {Robert C.~Kennicutt}, \citenamefont {Ford},
  \citenamefont {Graham}, \citenamefont {Huchra}, \citenamefont {Hughes},
  \citenamefont {Illingworth}, \citenamefont {Macri},\ and\ \citenamefont
  {Stetson}}]{FMG+01}%
  \BibitemOpen
  \bibfield  {author} {\bibinfo {author} {\bibfnamefont {W.~L.}\ \bibnamefont
  {Freedman}}, \bibinfo {author} {\bibfnamefont {B.~F.}\ \bibnamefont
  {Madore}}, \bibinfo {author} {\bibfnamefont {B.~K.}\ \bibnamefont {Gibson}},
  \bibinfo {author} {\bibfnamefont {L.}~\bibnamefont {Ferrarese}}, \bibinfo
  {author} {\bibfnamefont {D.~D.}\ \bibnamefont {Kelson}}, \bibinfo {author}
  {\bibfnamefont {S.}~\bibnamefont {Sakai}}, \bibinfo {author} {\bibfnamefont
  {J.~R.}\ \bibnamefont {Mould}}, \bibinfo {author} {\bibfnamefont
  {J.}~\bibnamefont {Robert C.~Kennicutt}}, \bibinfo {author} {\bibfnamefont
  {H.~C.}\ \bibnamefont {Ford}}, \bibinfo {author} {\bibfnamefont {J.~A.}\
  \bibnamefont {Graham}}, \bibinfo {author} {\bibfnamefont {J.~P.}\
  \bibnamefont {Huchra}}, \bibinfo {author} {\bibfnamefont {S.~M.~G.}\
  \bibnamefont {Hughes}}, \bibinfo {author} {\bibfnamefont {G.~D.}\
  \bibnamefont {Illingworth}}, \bibinfo {author} {\bibfnamefont {L.~M.}\
  \bibnamefont {Macri}}, \ and\ \bibinfo {author} {\bibfnamefont {P.~B.}\
  \bibnamefont {Stetson}},\ }\href@noop {} {\bibfield  {journal} {\bibinfo
  {journal} {The Astrophysical Journal}\ }\textbf {\bibinfo {volume} {553}},\
  \bibinfo {pages} {47} (\bibinfo {year} {2001})}\BibitemShut {NoStop}%
\bibitem [{\citenamefont {Ver~Steeg}\ and\ \citenamefont
  {Menicucci}(2009)}]{VerSteeg2009}%
  \BibitemOpen
  \bibfield  {author} {\bibinfo {author} {\bibfnamefont {G.}~\bibnamefont
  {Ver~Steeg}}\ and\ \bibinfo {author} {\bibfnamefont {N.~C.}\ \bibnamefont
  {Menicucci}},\ }\href {\doibase 10.1103/PhysRevD.79.044027} {\bibfield
  {journal} {\bibinfo  {journal} {Phys. Rev. D}\ }\textbf {\bibinfo {volume}
  {79}},\ \bibinfo {pages} {044027} (\bibinfo {year} {2009})}\BibitemShut
  {NoStop}%
\end{thebibliography}%

\end{document}